\documentclass[sigconf]{acmart}

\usepackage{xspace}
\usepackage{amsmath}

\usepackage{multirow}
\usepackage{subcaption}
\usepackage{graphicx}	
\usepackage{booktabs}
\usepackage{xcolor}
\usepackage{enumitem,kantlipsum}

\usepackage{balance}

\usepackage{colortbl}
\AtBeginDocument{%
  }

\copyrightyear{2024}
\acmYear{2024}
\setcopyright{rightsretained}
\acmConference[RecSys '24]{18th ACM Conference on Recommender Systems}{October 14--18, 2024}{Bari, Italy}

\acmBooktitle{18th ACM Conference on Recommender Systems (RecSys '24), October 14--18, 2024, Bari, Italy}\acmDOI{10.1145/3640457.3688187} \acmISBN{979-8-4007-0505-2/24/10}

\newcommand{\modelitemknn}{\texttt{ItemKNN}\xspace}

\newcommand{\modelbpr}{\texttt{BPR}\xspace}

\newcommand{\modelvae}{\texttt{MultiVAE}\xspace}

\newcommand{\modelpop}{\texttt{Pop}\xspace}

\newcommand{\modelneumf}{\texttt{NeuMF}\xspace}

\newcommand{\modellightgcn}{\texttt{LightGCN}\xspace}

\begin{document}

\title[Country Representation Dynamics Created by Feedback Loops in Music Recommender Systems]{Oh, Behave! Country Representation Dynamics Created by\\Feedback Loops in Music Recommender Systems}

\author{Oleg Lesota}
\email{oleg.lesota@jku.at}
\orcid{0000-0002-8321-6565}
\affiliation{%
  \institution{Johannes Kepler University Linz and \\ Linz Institute of Technology}
  \city{Linz}
  \country{Austria}
}

\author{Jonas Geiger}
\email{jonasgeiger@outlook.de}
\orcid{0009-0001-6426-6858}
\affiliation{%
  \institution{Johannes Kepler University Linz and \\ Linz Institute of Technology}
  \city{Linz}
  \country{Austria}
}

\author{Max Walder}
\email{max@gstat.eu}
\orcid{0009-0002-0644-801X}
\affiliation{%
  \institution{Johannes Kepler University Linz and \\ Linz Institute of Technology}
  \city{Linz}
  \country{Austria}
}

\author{Dominik Kowald}
\email{dkowald@know-center.at}
\orcid{0000-0003-3230-6234}
\affiliation{
  \institution{Know-Center GmbH and \\ Graz University of Technology}
  \city{Graz}
  \country{Austria}
}

\author{Markus Schedl}
\email{markus.schedl@jku.at}
\orcid{0000-0003-1706-3406}
\affiliation{
  \institution{Johannes Kepler University Linz and \\ Linz Institute of Technology}
  \city{Linz}
  \country{Austria}
}

\renewcommand{\shortauthors}{Lesota et al.}


\begin{abstract}
Recent work suggests that music recommender systems are prone to disproportionally frequent recommendations of music from countries more prominently represented in the training data, notably the US. However, it remains unclear to what extent feedback loops in music recommendation influence the dynamics of such imbalance. 
In this work, we investigate the dynamics of representation of local (i.e., country-specific) and US-produced music in user profiles and recommendations. 
To this end, we conduct a feedback loop simulation study using the LFM-2b dataset. 
The results suggest that most of the investigated recommendation models decrease the proportion of music from local artists in their recommendations. Furthermore, we find that models preserving average proportions of US and local music do not necessarily provide country-calibrated recommendations. 
We also look into popularity calibration and, surprisingly, find that the most popularity-calibrated model in our study (ItemKNN) provides the least country-calibrated recommendations.
In addition, users from less represented countries (e.g., Finland) are, in the long term, most affected  by the under-representation of their local music in recommendations. 
\end{abstract}

\ccsdesc[500]{Information systems~Recommender systems}
\ccsdesc[300]{Social and professional topics~Cultural characteristics}

\keywords{Music Recommendation, Feedback Loop, Demographic~Bias} 


\maketitle

\section{Introduction}
Recent research has demonstrated the effectiveness of long-term investigations in studying bias and representation dynamics induced by feedback loops in recommender systems and decision support systems~\cite{chaney2018algorithmic,DBLP:conf/cikm/MansouryAPMB20,d2020fairness,lacic2022drives,scher2023modelling}. For instance, Mansoury et al.~investigated the amplification of popularity bias~\cite{DBLP:conf/ecir/KowaldSL20,DBLP:conf/recsys/LesotaMRBKLS21-0,kowald2022popularity} by examining the average item popularity and aggregating diversity in movie recommendation lists across multiple iterations~\cite{DBLP:conf/cikm/MansouryAPMB20}. 
In~\citet{lacic2022drives}, the authors studied long-term dynamics in news recommender systems through an online study, measuring the impact of introducing a personalized recommendation algorithm on popularity bias compared to a purely popularity-based one. The results revealed that significant changes in users' news consumption behavior could be observed after the end of the study. 
Apart from popularity bias, previous research has investigated related phenomena, namely traces of globalization in music consumption patterns amplified by music recommender systems: Lesota et al.~identified a strong position of US-generated music (i.e., the majority group) in the consumption behavior of users from other countries~\cite{DBLP:conf/ismir/LesotaPBLRS22}. However, it is still unclear to what extent feedback loops in music recommendation influence country representation dynamics (of users and artists) in the long run.
To close this research gap, 
we study the following questions:
\begin{itemize}
    \item \textbf{RQ1}: How do different recommender systems affect representation of local and US-produced music in recommendations and user profiles in the long term?
    \item \textbf{RQ2}: How do effects of feedback loops vary across different countries? Do different recommender systems treat individual countries differently?
\end{itemize}

To the best of our knowledge this work is the first to (1) investigate the impact recommender systems can have on the representation of music from different countries in the \textit{long term}, and (2) describe the impact in terms of 
the degree of mismatch between country label distributions over the tracks in user consumption histories and their corresponding recommendations (\textit{miscalibration of country distributions}).

\section{Methodology}
We address the research questions by conducting an offline simulation study, following the general procedure introduced by~\citet{DBLP:conf/cikm/MansouryAPMB20}. 
We simulate user interaction with a recommendation model over a long period of time via a feedback loop. At each iteration of the loop, the model produces recommendations. 
One of the recommended items per user\footnote{We simulate consumption of only one item per user per iteration to avoid the situation of ``running out of item supply'', which may occur in our setting when items already consumed by a user can no longer be recommended to them.} is added to the respective user's interaction history (simulating item consumption). After that, the model is retrained on the data enriched with the new interactions of all users and then the process is repeated multiple times. The recommendations and simulated user profiles at each iteration are evaluated regarding the representation of local and US music.

\vspace{1mm} \noindent \textbf{Simulation procedure.} At each iteration, the following steps are performed for the recommendation algorithm under investigation:

\begin{enumerate}[wide, labelwidth=!, labelindent=0pt]
    \item We randomly split the input dataset into a training set $(75\%)$, validation set $(20\%)$ and test set $(5\%)$. 
    \item We train the recommendation model under investigation on the training set and use 
    NDCG@k as selection criterion on the validation set. 
    After the training process has terminated, the final model is used to produce $k$ recommendations for each user. Items already seen by the user are excluded. 
    \item From the recommended items $\{t\}$ generated by the model, one is added to the user profile based on an acceptance probability concept adapted from~\cite{DBLP:conf/cikm/MansouryAPMB20}. For each user and their list of $k$ recommended items ($R$), the probability is expressed as
    $prob(t|R) = \frac{e^{\alpha * rank_t}}{\sum_{j=1}^{k} e^{\alpha * j}}$
    where $rank_t$ is the rank of the item $t$ in the sorted list $R$ 
    and $\alpha$ is a hyperparameter $<0$.
    \item After an item has been selected for all users, a new dataset is created combining the input dataset with the new (simulated) interactions. 
    The obtained dataset is then used for the next simulation iteration as the new input dataset. 
\end{enumerate}

\vspace{1mm} \noindent \textbf{Evaluation.} The following indicators and metrics are used to evaluate country representation in user recommendations and profiles. 
\textit{US and local proportions}  show how many items in the profile or recommendations of user $u$ originate from the US or the country of the user. The local proportion $p_{u}^{local}$ for a given set of interactions or recommendations $I_u$ of a user is defined as $p_{u}^{local} = \frac{|I_{u}^{local}|}{|I_u|}$, where $I_{u}^{local}$ is the set of all tracks where user country and track country coincide. The US proportion $p_{u}^{US}$ is defined analogously. The proportions are defined on the user level and averaged over users. We analyze the difference in these proportions between the original, unmodified dataset and the state of the recommendations or user profiles at iteration $i$.
\textit{Popularity and country miscalibration} are used to show how much the user profiles deviate from their original states after $n$ iterations of the simulation. Following the assumption that the users prefer calibrated recommendations \cite{DBLP:conf/recsys/Steck18}, we compute the miscalibration of their profiles with regard to the music track attributes of country and popularity as Jensen–Shannon divergence~($JSD$), see Equation~\ref{eq:jsd}.

\begin{equation} \label{eq:jsd}
\scalebox{0.85}{$
\begin{aligned}
    JSD(H_u, H^*_u) = \frac{1}{2}\sum_{c}{H_u(c) \log_2\frac{2H_u(c)}{H_u(c) + H^*_u(c)}} + \\
    \frac{1}{2}\sum_{c}{H^*_u(c) \log_2\frac{2H^*_u(c)}{H_u(c) + H^*_u(c)}}
\end{aligned}$}
\end{equation}

$H_u$ is the distribution of a track attribute over unmodified interaction history of user $u$ and $H^*_u$ is the same for the interaction history after $n$ iterations. $H(c)$ is the (probability) value in the distribution~$H$, corresponding to the track attribute value $c$. In case of popularity, $c \subset [HighPop, MidPop, LowPop]$ defined as in \cite{um/AbdollahpouriMB21, DBLP:conf/recsys/LesotaBWMLRS22}. For country $c \subset [local, US, other]$. 
The per-user $JSD$ scores are averaged to represent miscalibration affecting a user group.
\section{Experimental setup}
To foster reproducibility, we make the source code, data sample and other materials publicly available on GitHub\footnote{https://github.com/hcai-mms/FeedbackLoops4RecSys}.

\vspace{0.1cm}
\noindent \textbf{Dataset.} 
We conduct our experiments on the LFM-2b dataset \cite{DBLP:conf/chiir/SchedlBLPPR22}. It contains listening events from the online music aggregator platform Last.fm.\footnote{\texttt{https://www.last.fm}}
First, we apply the same pre-processing and augmentation operations as \citet{DBLP:conf/ismir/LesotaPBLRS22}. Specifically, we extend the dataset with information about the country of origin of artists from Musicbrainz\footnote{\texttt{https://musicbrainz.org}} to attribute a country to each music track. 
We then filter (1) all tracks listened to only once, and (2) all interactions for which no country information about the user or track is available. We consider interactions from the years 2018-2019, the two most recent full years in the dataset.
Additionally, due to computational limitations, we only work with a sub-sample of this filtered dataset. We randomly sample 100K tracks, ensuring that the resulting data sample is 5-core filtered. This results in a dataset of $\sim$2290K interactions between $\sim$100K tracks and $\sim$12K users. Detailed statistics are reported in Table \ref{tab:dataset_statistics}.

\begin{table}
    \centering
    \caption{Basic statistics of our LFM-2b dataset sample.}
    \label{tab:dataset_statistics}
    \scalebox{0.85}{
    \begin{tabular}{l|r|r|r|r|r} 
        \toprule
        & Tracks & \multicolumn{2}{c|}{Track Interactions} & Users & User \\
        \cline{3-4}
        & & Total & Average & & Interactions\\
        \midrule
        US & 39,614 & 1,040,360 & 26.26  & 1,582 & 323,072 \\
        UK & 15,522 & 422,225 & 27.20 & 823 & 171,469 \\
        DE & 6,793 & 107,832 & 15.87 & 805 & 158,642 \\
        SE & 4,519 & 107,491 & 23.79 & 320 & 60,993 \\
        CA & 3,754 & 95,343 & 25.40 & 217 & 47,490 \\
        FR & 2,800 & 56,241 & 20.09 & 254 & 52,850 \\
        AU & 2,346 & 53,701 & 22.89 & 193 & 40,767 \\
        FI & 2,260 & 45,709 & 20.23 & 420 & 78,819 \\
        NO & 1,765 & 36,769 & 20.83 & 208 & 40,545 \\
        BR & 2,236 & 35,964 & 16.08 & 1,064 & 205,093 \\
        NL & 1,738 & 32,035 & 18.43 & 375 & 89,546 \\
        PL & 1,709 & 27,116 & 15.87 & 1,040 & 195,296 \\
        RU & 1,888 & 24,086 & 12.76 & 1,162 & 187,876 \\
        JP & 1,796 & 21,818 & 12.15 & 101 & 14,411 \\
        IT & 1,506 & 21,273 & 14.13 & 222 & 37,421 \\
        other & 9,651 & 159,769 & 16.55 & 2,990 & 583,442 \\
        \midrule
        Total & 99,897 & 2,287,732 & 22.90 & 11,776 & 2,287,732 \\
        \bottomrule 
    \end{tabular}
    }
\end{table}

\vspace{1mm} \noindent \textbf{Models.} 
We perform a feedback loop simulation with $n = 100$ iterations using the method described above for each of the following six models: \modelbpr\cite{Rendle2009BPR}, \modelitemknn\cite{Deshpande2004ItemKNN}, \modellightgcn\cite{He2020LightGCN}, \modelvae\cite{Xu2021MultiVAE}, \modelneumf\cite{He2017NeuMF}, and \modelpop. \modelitemknn is a traditional approach that assigns recommendation scores to items based on how similar an item is to those already consumed by a user, depending on the existing interactions of other users. 
Bayesian Personalized Ranking (\modelbpr) is a matrix factorization~\cite{muellner2021robustness} approach using a personalized loss, which enforces explicit ranking between pairs of items.
\modellightgcn is a graph-based approach employing a Graph Convolution Network, where the reconstruction of the interaction matrix also considers the embeddings of neighboring users and items on the user-item interaction graph. Neural Matrix Factorization (\modelneumf) employs a multi-layer perceptron, allowing it to learn a custom matching function. 
\modelvae is a variational autoencoder that learns a latent user representation from users' interaction vectors, and predicts the distribution of the relevance score over all items. 
The \modelpop model recommends the most popular unseen items to each user, where popularity is computed as frequency of interactions over all users. It serves as a baseline model against which the more complex models are compared.

\vspace{1mm} \noindent \textbf{Training and evaluation procedure.}
We run  our experiments using the Recbole library \cite{Zhao2021Recbole} and its default settings for each model. We train for a maximum of 200 epochs, with early stopping applied after not observing any improvement in terms of NDCG@10 for 5~subsequent epochs. For the simulated acceptance of new items, we consider the items with top $k=10$ scores for each user. The acceptance probability is calculated with $\alpha=-0.1$. Following \cite{DBLP:conf/ismir/LesotaPBLRS22}, when analyzing the results for individual countries we concentrate on ones that are represented by at least 100 users and 1000 tracks.\footnote{We abbreviate the countries as follows.  US: United States, UK: United Kingdom, DE: Germany, SE: Sweden, CA: Canada, FR: France, AU: Australia, FI: Finland, BR: Brazil, RU: Russia, JP: Japan, NO: Norway, PL: Poland, NL: Netherlands, IT: Italy}

\section{Results}
\begin{figure} 
    \centering
    \scalebox{0.82}{
    \includegraphics[width=1\linewidth]{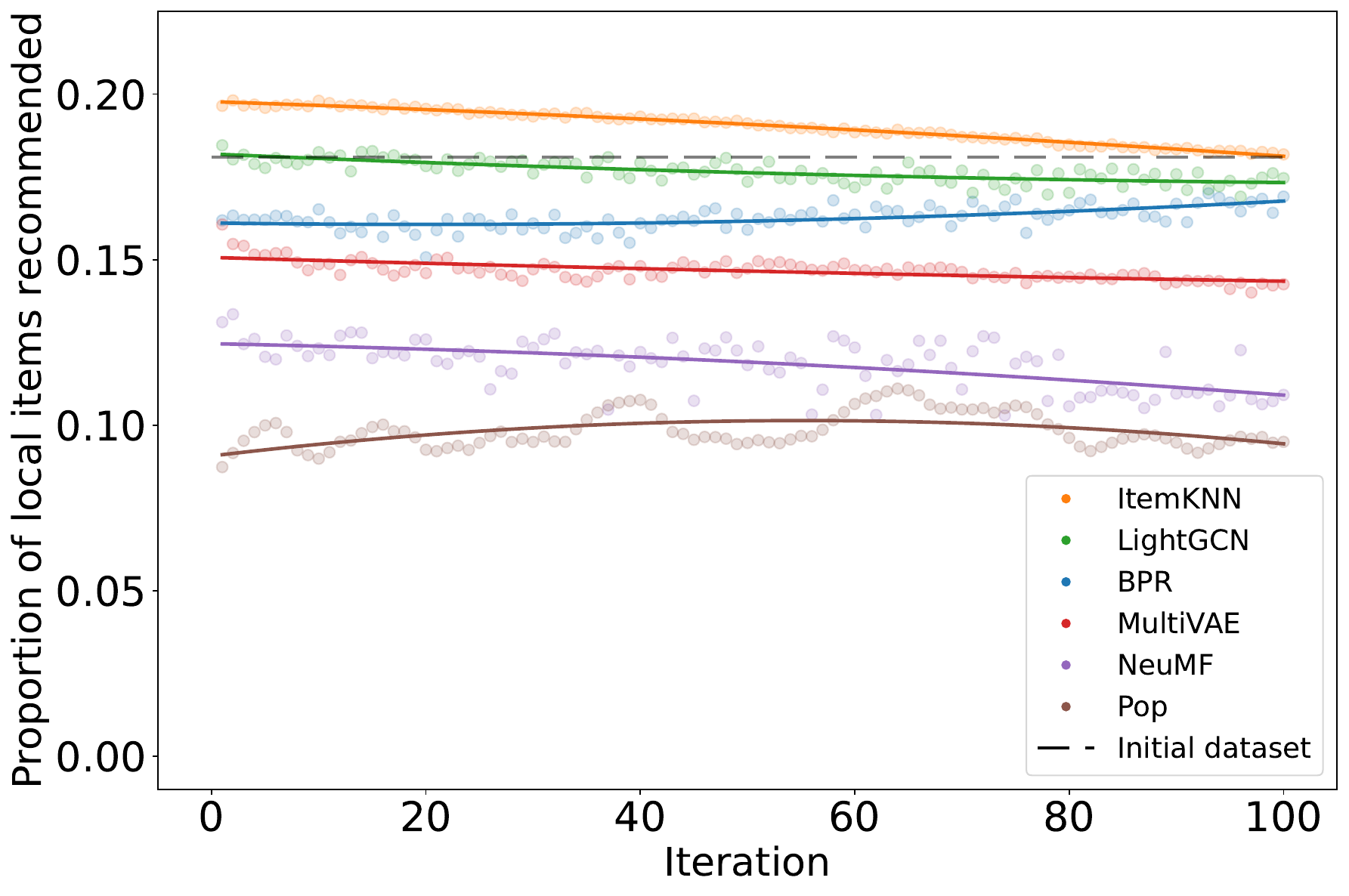}}
    \caption{Proportion of local items recommended by different algorithms. The dashed line shows the average consumption of local music before the simulation.}
    \label{fig:local-per-algorithm}
    \Description[Fully described in the Results section]{Fully described in the Results section}
\end{figure}

\begin{figure}
    \centering
    \includegraphics[width=1\linewidth]{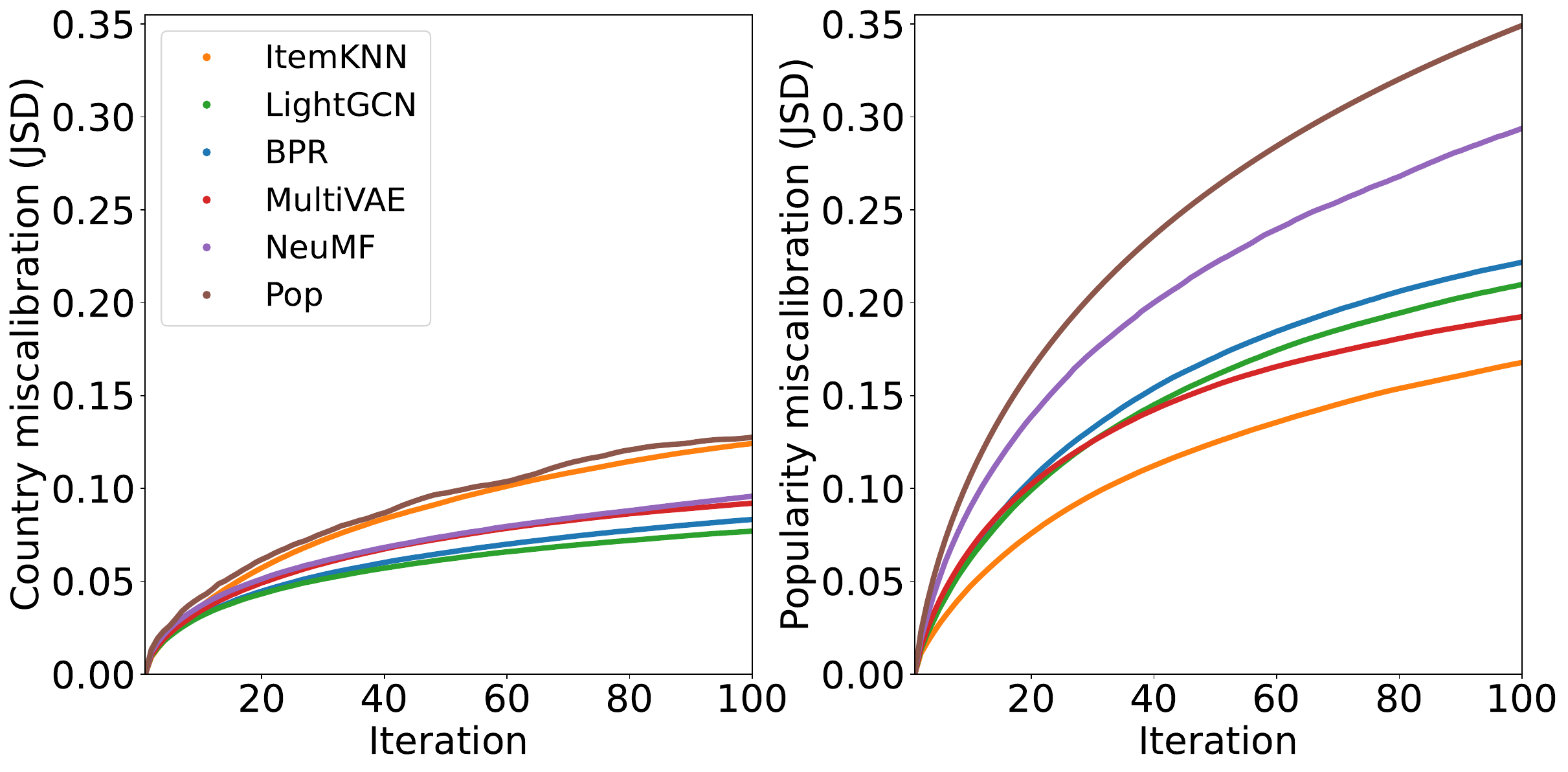}
    \caption{Miscalibration between the interaction history at iteration $i$ and the initial interaction history in terms of country (left) and popularity (right). Measured as $JSD$.\vspace{-3mm}}
    \label{fig:country-pop-jsd-per-algorithm}
    \Description[Most country miscalibrated ItemKNN and Pop (0.12), least LightGCN (0.07). Most pop miscalibrated Pop (0.35), least (0.15).]{From most to least miscalirated recommenders at iteration 100. Country miscalibration: ItemKNN and Pop (JSD of about 0.12), NeuMF, MultiVAE, BPR, LightGCN (JSD of about 0.07). Popularity miscalibration: Pop (JSD of about 0.35), NeuMF, BPR, LightGCN, MultiVAE, ItemKNN (JSD of about 0.15).}
\end{figure}

We summarize the main results concerning \textbf{RQ1} in Table~\ref{tab:whole_population} and Figure~\ref{fig:local-per-algorithm}. Table~\ref{tab:whole_population} presents the simulated impact the six investigated recommenders made on user recommendations and user profiles. The rows $Rec_{local}$ and $Rec_{US}$ show how different the proportions of local and US-produced music recommended at iteration $100$ are to the respective proportions in the original user profiles in percent. Similarly, the rows $Prof_{local}$ and $Prof_{US}$ show the difference between user profiles before and after the simulation. For example, the value of $+17.0^{*}$ for $Rec_{US}$ and \modelneumf means that at the last iteration of the simulation, the average proportion of US music recommended by this model is $17\%$ higher than the proportion of the US music consumed by the users before the simulation. 
Statistically significant\footnote{As determined by a two-sided T-Test with the Bonferroni-corrected threshold of $0.05/12$ as each initial proportion is compared to ones produced by each of 6 algorithms both in the recommendations and in user profiles.} inconsistencies are marked with $^{*}$. $JSD_{Prof}$ shows the miscalibration in terms of country distribution between user profiles before and after the simulation. $nDCG_1$ shows model accuracy evaluation at iteration $1$, verifying that at the initial state, every model has shown adequate performance for the dataset (similar 
results were achieved on other sub-samples of LFM-2b \cite{DBLP:conf/recsys/MelchiorreZS20, DBLP:conf/recsys/LesotaBWMLRS22}). 
In Figure~\ref{fig:local-per-algorithm} every point corresponds to the average proportion of local music in recommendations across all users per algorithm per iteration. The dashed line ($0.18$) shows the proportion of consumed local music in the initial dataset.
We approach \textbf{RQ2} analyzing Table~\ref{tab:c_jsd_prof} and Table~\ref{tab:prop_per_country}. The former shows miscalibration between country distributions over tracks in user profiles before and after simulation for each model and each of the considered countries, expressed as $JSD$. The countries in the table are arranged in the order of decreasing number of respective tracks. The highest and lowest values of miscalibration among non-baseline algorithms for each country are put in bold and underlined, respectively. Table~\ref{tab:prop_per_country} shows the mismatch in average proportions of local and US-produced music in user profiles before and after the simulation in percent. 
We make additional observations regarding popularity and country calibration using Figure~\ref{fig:country-pop-jsd-per-algorithm} showing the progression of miscalibration between the simulated user profiles at each iteration and their original state. 

\begin{table}
    \centering
    \caption{Proportion of $local$ and $US$-produced music in recommendations ($Rec_{*}$) and simulated user profiles ($Prof_{*}$) after iteration $100$ compared to the respective proportions in the original user profiles (difference in $\%$). Significant changes are marked with $^{*}$.} 
    \label{tab:whole_population}
    \scalebox{0.80}{
    \begin{tabular}{l|c|c|c|c|c|c} \toprule
         & \modelpop & \modelitemknn & \modelbpr & \modelneumf & \modelvae & \modellightgcn \\ \midrule
        $Rec_{local}$ & $-47.5^{*}$ & $+0.4$ & $-6.6^{*}$ & $-39.7^{*}$ & $-21.2^{*}$ & $-3.5$ \\
        $Rec_{US}$ & $+15.6^{*}$ & $+4.8^{*}$ & $+3.5^{*}$ & $+17.0^{*}$ & $+7.5^{*}$ & $+0.5$ \\ \midrule
        $Prof_{local}$ & $-22.5^{*}$ & $+2.1$ & $-7.0^{*}$ & $-19.5^{*}$ & $-10.5^{*}$ & $-3.6$ \\
        $Prof_{US}$ & $+19.5^{*}$ & $+2.2^{*}$ & $+3.7^{*}$ & $+7.6^{*}$ & $+2.6^{*}$ & $+0.8$ \\\midrule \midrule
        $JSD_{Prof}$ & $0.13$ & $0.12$ & $0.08$ & $0.10$ & $0.09$ & $0.08$  \\ \midrule 
        $nDCG_1$ & $0.03$ & $0.26$ & $0.13$ & $0.08$ & $0.12$ & $0.14$  \\ \bottomrule 
    \end{tabular}
    }
\end{table}

\subsection{RQ1: Representation of Local and US-produced Music in the Long Term}
We observe that the recommendations produced by all models, except for \modellightgcn, at iteration $100$ significantly differ from the initial user profiles in terms of average proportions of US and local music (Table~\ref{tab:whole_population}); with an overall trend of increasing average proportions for US music and decreasing for local music (except for \modelitemknn) in recommendations. Figure~\ref{fig:local-per-algorithm} shows that at iteration~1 all models, except for \modellightgcn, on average suggest a proportion of local tracks inconsistent with the initial dataset. The proportions of local tracks deviate further away in recommendations by \modelvae and \modelneumf and potentially converge to the initial level for \modelitemknn and \modelbpr. \modelneumf shows the highest deviation, comparable to \modelpop. The user profiles' shift from their original state corresponds to the discussed shift in the recommendations (Table~\ref{tab:whole_population}). Additionally, \modellightgcn causes the least country-miscalibrated ($JSD_{Prof}$) user profiles on average, while the highest miscalibration in terms of country is caused by \modelitemknn. 

\vspace{1mm} \noindent \textbf{Main findings.}  
We conclude that recommender systems tend to over-represent US and under-represent local music. However, some models (\modelitemknn, \modelbpr, \modellightgcn) may converge to the original levels of local representation in the long term. The overall algorithmic impact varies from the most calibrated \modellightgcn that also preserves the initial proportions, to the most miscalibrated \modelitemknn still preserving the initial proportions, and notably miscalibrated \modelneumf significantly affecting both country proportions.


\begin{table*}
    \centering
    \caption{Miscalibration between three-bin country distributions over the original user profiles and the profiles after $100$ iterations (measured as $JSD$). Higher values indicate higher inconsistency. Darker background indicates higher miscalibration, normalized per algorithm.
    \vspace{-3mm}} \label{tab:c_jsd_prof}
    \scalebox{0.82}{
    \begin{tabular}{l|c|c|c|c|c|c|c|c|c|c|c|c|c|c|c||c}
    \toprule
         & US & UK & DE & SE & CA & FR & AU & FI & NO & BR & NL & PL & RU & JP & IT & all \\ \midrule
        \modelpop & \cellcolor{teal!4} $0.073$ & \cellcolor{teal!24} $0.126$ & \cellcolor{teal!40} $0.146$ & \cellcolor{teal!28} $0.133$ & \cellcolor{teal!16} $0.108$ & \cellcolor{teal!36} $0.144$ & \cellcolor{teal!12} $0.101$ & \cellcolor{teal!56} $0.186$ & \cellcolor{teal!16} $0.108$ & \cellcolor{teal!28} $0.133$ & \cellcolor{teal!8} $0.096$ & \cellcolor{teal!48} $0.16$ & \cellcolor{teal!52} $0.163$ & \cellcolor{teal!60} $0.196$ & \cellcolor{teal!44} $0.153$ & 0.128 \\ \midrule
        \modelitemknn & \cellcolor{teal!4} $\textbf{0.084}$ & \cellcolor{teal!20} $\textbf{0.123}$ & \cellcolor{teal!40} $\textbf{0.139}$ & \cellcolor{teal!36} $\textbf{0.137}$ & \cellcolor{teal!12} $\textbf{0.106}$ & \cellcolor{teal!28} $\textbf{0.128}$ & \cellcolor{teal!16} $\textbf{0.109}$ & \cellcolor{teal!48} $\textbf{0.152}$ & \cellcolor{teal!24} $\textbf{0.127}$ & \cellcolor{teal!32} $\textbf{0.134}$ & \cellcolor{teal!8} $\textbf{0.094}$ & \cellcolor{teal!52} $\textbf{0.155}$ & \cellcolor{teal!44} $\textbf{0.149}$ & \cellcolor{teal!56} $0.166$ & \cellcolor{teal!60} $\textbf{0.174}$ & 0.124 \\
        \modelbpr & \cellcolor{teal!4} $0.054$ & \cellcolor{teal!36} $0.093$ & \cellcolor{teal!28} $0.089$ & \cellcolor{teal!32} $0.09$ & \cellcolor{teal!20} $0.085$ & \cellcolor{teal!40} $0.095$ & \cellcolor{teal!24} $\underline{0.088}$ & \cellcolor{teal!48} $0.103$ & \cellcolor{teal!16} $\underline{0.084}$ & \cellcolor{teal!12} $0.083$ & \cellcolor{teal!8} $0.07$ & \cellcolor{teal!48} $0.103$ & \cellcolor{teal!44} $0.098$ & \cellcolor{teal!60} $0.146$ & \cellcolor{teal!56} $\underline{0.114}$ & 0.083 \\
        \modelneumf & \cellcolor{teal!4} $0.058$ & \cellcolor{teal!24} $0.099$ & \cellcolor{teal!40} $0.114$ & \cellcolor{teal!32} $0.104$ & \cellcolor{teal!16} $0.09$ & \cellcolor{teal!36} $0.111$ & \cellcolor{teal!20} $0.093$ & \cellcolor{teal!52} $0.126$ & \cellcolor{teal!12} $0.087$ & \cellcolor{teal!24} $0.099$ & \cellcolor{teal!8} $0.074$ & \cellcolor{teal!48} $0.121$ & \cellcolor{teal!44} $0.115$ & \cellcolor{teal!60} $\textbf{0.179}$ & \cellcolor{teal!56} $0.135$ & 0.096 \\
        \modelvae & \cellcolor{teal!4} $0.062$ & \cellcolor{teal!24} $0.098$ & \cellcolor{teal!40} $0.101$ & \cellcolor{teal!28} $0.099$ & \cellcolor{teal!16} $0.092$ & \cellcolor{teal!32} $0.1$ & \cellcolor{teal!20} $0.094$ & \cellcolor{teal!52} $0.114$ & \cellcolor{teal!12} $0.091$ & \cellcolor{teal!32} $0.1$ & \cellcolor{teal!8} $0.074$ & \cellcolor{teal!48} $0.111$ & \cellcolor{teal!44} $0.108$ & \cellcolor{teal!56} $0.132$ & \cellcolor{teal!60} $0.136$ & 0.092 \\
        \modellightgcn & \cellcolor{teal!4} $\underline{0.053}$ & \cellcolor{teal!44} $\underline{0.089}$ & \cellcolor{teal!16} $\underline{0.079}$ & \cellcolor{teal!24} $\underline{0.081}$ & \cellcolor{teal!20} $\underline{0.08}$ & \cellcolor{teal!32} $\underline{0.086}$ & \cellcolor{teal!48} $0.091$ & \cellcolor{teal!40} $\underline{0.088}$ & \cellcolor{teal!28} $0.084$ & \cellcolor{teal!12} $\underline{0.069}$ & \cellcolor{teal!8} $\underline{0.068}$ & \cellcolor{teal!52} $\underline{0.093}$ & \cellcolor{teal!32} $\underline{0.086}$ & \cellcolor{teal!56} $\underline{0.099}$ & \cellcolor{teal!60} $0.13$ & 0.077 \\
    \bottomrule
    \end{tabular}}
\end{table*}

\begin{table*}
    \centering
    \caption{Deviations in the average proportions of local and US music in user profiles at iteration $100$ from the respective average proportions in the original user profiles before simulation (in percent). Statistically significant deviations are marked with $^{*}$. The highest value for each column for each indicator is depicted in bold, the lowest is \underline{underlined}. \vspace{-3mm}} \label{tab:prop_per_country}
        \scalebox{0.78}{
        \begin{tabular}{l|c|c|c|c|c|c|c|c|c|c|c|c|c|c|c||c}
        \toprule
 & US & UK & DE & SE & CA & FR & AU & FI & NO & BR & NL & PL & RU & JP & IT & all \\ \midrule
\multicolumn{17}{c}{local proportion in user profiles} \\  \midrule
\modelpop & $-1.2$ & $-4.2$ & $-41.6^{*}$ & $-40.8^{*}$ & $-31.0^{*}$ & $-42.2^{*}$ & $-31.2^{*}$ & $-53.4^{*}$ & $-47.3^{*}$ & $-48.6^{*}$ & $-44.1^{*}$ & $-54.4^{*}$ & $-57.7^{*}$ & $-67.3^{*}$ & $-58.3^{*}$ & $-22.5^{*}$ \\ \midrule
\modelitemknn & $\textbf{-1.5}$ & $\underline{-15.0}^{*}$ & $\textbf{+5.6}$ & $\textbf{+2.4}$ & $\textbf{-18.3}$ & $-20.1$ & $\textbf{-25.6}$ & $\textbf{+16.0}$ & $\textbf{-1.7}$ & $\textbf{+7.2}$ & $\textbf{-19.6}$ & $\textbf{+28.5}^{*}$ & $\textbf{+10.7}$ & $\textbf{+18.7}$ & $\textbf{+49.6}$ & $\textbf{+2.1}$ \\
\modelbpr & $-2.3$ & $-7.6^{*}$ & $-8.8$ & $-17.9^{*}$ & $\underline{-23.7}^{*}$ & $-31.3^{*}$ & $-31.3^{*}$ & $+3.4$ & $-39.2^{*}$ & $+3.8$ & $-35.1^{*}$ & $-1.2$ & $-19.1^{*}$ & $-40.3$ & $-5.6$ & $-7.0^{*}$ \\
\modelneumf & $-3.3^{*}$ & $\textbf{-6.1}$ & $\underline{-33.6}^{*}$ & $\underline{-31.4}^{*}$ & $-23.1^{*}$ & $\underline{-40.2}^{*}$ & $\underline{-33.8}^{*}$ & $\underline{-35.0}^{*}$ & $\underline{-39.9}^{*}$ & $\underline{-29.5}^{*}$ & $\underline{-37.3}^{*}$ & $\underline{-46.7}^{*}$ & $\underline{-49.0}^{*}$ & $\underline{-66.5}^{*}$ & $\underline{-56.6}^{*}$ & $\underline{-19.5}^{*}$ \\
\modelvae & $\underline{-3.5}^{*}$ & $-7.5^{*}$ & $-7.0$ & $-19.0^{*}$ & $-23.4^{*}$ & $-34.8^{*}$ & $-29.8^{*}$ & $-19.0$ & $-38.7^{*}$ & $-0.4$ & $-36.3^{*}$ & $-23.6^{*}$ & $-12.7$ & $-18.5$ & $-53.1^{*}$ & $-10.5^{*}$ \\
\modellightgcn & $-3.1$ & $-10.3^{*}$ & $-3.4$ & $-1.7$ & $-21.3^{*}$ & $\textbf{-3.7}$ & $-29.5^{*}$ & $-0.8$ & $-35.4^{*}$ & $-3.5$ & $-31.5^{*}$ & $+12.3$ & $-12.3$ & $+10.0$ & $+48.6^{*}$ & $-3.6$ \\\midrule
\multicolumn{17}{c}{US proportion in user profiles} \\ \midrule
\modelpop & $-1.2$ & $+14.9^{*}$ & $+28.8^{*}$ & $+23.8^{*}$ & $+2.3$ & $+30.9^{*}$ & $+8.4^{*}$ & $+44.5^{*}$ & $+17.5^{*}$ & $+17.4^{*}$ & $+16.5^{*}$ & $+34.8^{*}$ & $+36.9^{*}$ & $+28.8^{*}$ & $+24.1^{*}$ & $+19.5^{*}$ \\ \midrule
\modelitemknn & $\textbf{-1.5}$ & $\textbf{+8.4}^{*}$ & $+2.9$ & $\underline{-0.7}$ & $-0.4$ & $+7.7$ & $\textbf{+4.5}$ & $\underline{-6.3}$ & $+1.5$ & $\underline{-0.4}$ & $+5.7$ & $+3.8$ & $+6.1$ & $\underline{-10.8}$ & $-1.9$ & $+2.2^{*}$ \\
\modelbpr & $-2.3$ & $+7.3^{*}$ & $+4.3$ & $+6.0$ & $\textbf{+0.3}$ & $+9.8$ & $+3.2$ & $0.0$ & $\textbf{+7.0}$ & $+2.5$ & $+4.8$ & $+6.9^{*}$ & $+7.0^{*}$ & $-0.4$ & $+3.1$ & $+3.7^{*}$ \\
\modelneumf & $-3.3^{*}$ & $+8.3^{*}$ & $\textbf{+13.6}^{*}$ & $\textbf{+10.6}^{*}$ & $\underline{-1.2}$ & $\textbf{+15.4}^{*}$ & $+2.9$ & $\textbf{+15.6}^{*}$ & $+6.5$ & $\textbf{+9.7}^{*}$ & $\textbf{+6.7}$ & $\textbf{+14.2}^{*}$ & $\textbf{+11.8}^{*}$ & $\textbf{+17.3}^{*}$ & $\textbf{+13.0}^{*}$ & $\textbf{+7.6}^{*}$ \\
\modelvae & $\underline{-3.5}^{*}$ & $\underline{+5.2}$ & $+2.4$ & $+3.4$ & $+0.1$ & $+6.0$ & $\underline{+2.6}$ & $+4.4$ & $+5.4$ & $+2.8$ & $+4.4$ & $+6.4^{*}$ & $+3.0$ & $+0.7$ & $+11.0^{*}$ & $+2.6^{*}$ \\
\modellightgcn & $-3.1$ & $+5.2$ & $\underline{+0.7}$ & $+0.2$ & $-0.9$ & $\underline{+3.0}$ & $+2.7$ & $-2.6$ & $\underline{+0.4}$ & $+1.9$ & $\underline{+3.4}$ & $\underline{+3.2}$ & $\underline{+2.9}$ & $-5.6$ & $\underline{-8.2}$ & $\underline{+0.8}$ \\
\bottomrule
    \end{tabular}}
\end{table*}

\subsection{RQ2: Effects of Feedback Loops Across Different Countries}
Table~\ref{tab:c_jsd_prof} demonstrates that the majority of algorithms treat users from FI, PL, RU, JP, and IT with the highest country miscalibration. The latter four countries are among the ones with the fewest tracks and the least average interactions per track. On the contrary, FI shows above median number of tracks and average track popularity, more than BR, which however is less affected by the miscalibration. Brazilian users are more frequently represented in the training data, which could be the reason for more calibrated recommendations. Among the countries experiencing lower country miscalibration are US, NL, CA, AU, and NO. The latter four are among the least represented. Notably, UK features more tracks and higher average track popularity than the four, while suffering higher miscalibration. Previous work \cite{DBLP:conf/ismir/LesotaPBLRS22} suggests that CA, NL, AU, and NO consume above average US-produced and below average of domestic music, unlike UK or DE.
\textit{The results show that feedback loops affect countries differently depending on multiple factors from country representation in the training data to consumption habits of respective users. More research is needed to determine clearer dependencies.}

Table~\ref{tab:prop_per_country} shows that \modelitemknn significantly affected average proportions of local or US music only for UK and PL. \modelneumf, however, significantly affected at least one of the proportions for all considered countries. Except for \modelneumf, all non-baseline models over-represent US-music for a limited number of countries, while the local music is under-represented for more combinations of country and model. This leads us to believe that the local music is gradually replaced by tracks from countries beyond the US and thus the progressive under-representation of local music is not necessarily caused by over-representation of music from one particular country (e.g. US), but rather by its own initial low representation.

\vspace{1mm} \noindent \textbf{Main findings.}  
The list of significantly affected countries varies between algorithms. Overall on per-country level most changes are in local rather than US proportions.

\subsection{Notes on Calibration}
Comparing the progression of miscalibration of user profiles from their original state in terms of country and popularity distributions (Figure~\ref{fig:country-pop-jsd-per-algorithm}), we observe that in both cases at iteration 100 the algorithms can be ranked according to the degree of miscalibration. Notably the rankings for popularity and country miscalibration do not match, for instance while \modelitemknn shows the lowest popularity miscalibration it results in the most country-miscalibrated user profiles. 

\vspace{1mm} \noindent \textbf{Main findings.}  
This suggests that while country miscalibration is likely connected to popularity bias, the connection is not direct, as the least popularity-biased model provides the most country-miscalibrated results.

\section{Limitations and Discussion}
In this section, we describe the limitations of our work and discuss our main findings mentioned in the previous section.

\vspace{1mm} \noindent \textbf{Country and culture.} 
In this work, we investigate the dynamics of country representation in user profiles and recommendations. It is worth pointing out that in our study, the country labels are assigned to tracks based on ``… the area with which an artist is primarily identified with. It is often, but not always, its birth/formation country…''\footnote{https://musicbrainz.org/doc/Artist\#Area}. This means that the country label does not necessarily reflect the language of the track, or to which extent the track is representative of the culture of the corresponding country. This limits the scope of our findings to the representation of artists from certain countries, which, while valuable, can be only a weak indicator of cultural representation. At the same time, formalizing the cultural signature of a music piece is a non-trivial task. The same could be said about assigning a language label to a track when it comes to classical, instrumental, electronic music or music preformed in dead languages. With this paper we strive to ignite the discussion about representation in recommender systems in terms of culture, country, language and other attributes for all stakeholders and suggest one possible approach to studying it.

\vspace{1mm} \noindent \textbf{Choice model.}  
When investigating feedback loops, we have to select a user choice model as a rule for simulating user interaction with recommendation lists. In our case, at each iteration the model randomly ``consumes'' one item per user with slight preference towards higher ranking items. In the real world scenario, users may consume more or fewer items per recommendation and are also able to consume items that were not recommended to them. In addition, every user is likely to have their own unique choice model(s). These discrepancies may affect the way the effects we observe in this paper would manifest for actual users. We see the study at hand as a reference point for further investigations considering alternative user choice models.

\vspace{1mm} \noindent \textbf{Confounding factors.}  
We conduct our experiments on a sample of the LFM-2b dataset collected from the Last.fm platform. Naturally, depending on how popular the platform is in different countries, they are represented in the data sample to a different extent in terms of both users and items. While, to an extent, representative of the real world, this, in combination with other factors, does not allow us to identify key factors affecting country representation in the recommendations. One approach to address this would be to conduct an ``ablation study'', a series of experiments on sub-samples balanced in terms of various attributes.

\section{Conclusion and Future Work}
The work at hand shows that in the context of a feedback loop, recommender systems tend to over-represent US and under-represent local music in their recommendations, potentially causing a shift in user preferences in the long run. 
We conclude that users whose countries are less represented in the data (e.g., Finland) are more likely to receive recommendations inconsistent with their original preferences in terms of artist country representation. However, higher representation does not always guarantee the opposite (e.g., Germany). We also find that \modelneumf affects the proportions of US and local music the most. At the same time, \modelitemknn provides least calibrated recommendations in terms of country representation while showing the highest calibration in terms of popularity. This suggests that country calibration does not necessarily follow popularity calibration. 

\vspace{1mm} \noindent \textbf{Future work.} 
We plan to investigate the connection between country-calibrated recommendations and popularity-calibrated recommendations. Other potential directions for future work include investigating alternative choice models and additional recommendation scenarios (repeated item consumption, sequential recommendation), defining causes of inconsistencies for every particular country, exploring various mitigation strategies as well as investigating the language (of a track and spoken by the user) and other cultural aspects of country representation.

\begin{acks}
    This research was funded in whole or in part by the Austrian Science Fund (FWF):  \url{https://doi.org/10.55776/P33526}, \url{https://doi.org/10.55776/DFH23}, \url{https://doi.org/10.55776/COE12}, \url{https://doi.org/10.55776/P36413}. 
    This research is funded by the Know-Center within the COMET — Competence Centers for Excellent Technologies Programme, funded by bmvit, bmdw, FFG, and SFG.
\end{acks}


\bibliographystyle{ACM-Reference-Format}
\balance
\bibliography{recsys24_feedback_loop}





\end{document}